\documentclass[12pt,preprint]{aastex}

\slugcomment{Not to appear in Nonlearned J., 45.}

\shorttitle{Wide Companion to GJ 282 AB}

\shortauthors{Poveda et al.}

\begin{document}


\title{G 112-29 (=NLTT 18149), a Very Wide Companion to GJ 282 AB with
a Common Proper Motion, Common Parallax, Common Radial Velocity and
Common Age}


\author{A. Poveda, Christine Allen,
R. Costero, J. Echevarr\'{i}a, and A. Hern\'{a}ndez-Alc\'{a}ntara,}

\affil{Instituto de Astronom\'{\i}a,
  Universidad Nacional Aut\'onoma de M\'exico, Apdo. Postal
  70--264, 04510, M\'exico. D. F., Mexico (chris@astroscu.unam.mx).
\email{chris@astroscu.unam.mx}}

\begin{abstract}
We have made a search for common proper motion (CPM) companions to
the wide binaries in the solar vicinity. We found that the binary GJ
282AB has a very distant CPM companion (NLTT 18149) at a separation
$s=1.09 \arcdeg$. Improved spectral types and radial velocities are
obtained, and ages determined for the three components.  The
Hipparcos trigonometric parallaxes and the new radial velocities and
ages turn out to be very similar for the three stars, and provide
strong evidence that they form a physical system. At a projected
separation of 55733AU from GJ 282AB, NLTT 18149 ranks among the
widest physical companions known.
\end{abstract}

\keywords{Binaries: visual --- Stars: individual (G112-29, GJ
282AB)}

\section{INTRODUCTION}
\label{sec:intro}

Our long-standing interest in the properties of very wide binaries
and their process of dissociation led us to search for additional
common proper motions (CPM) companions to the primaries of our
catalogue of wide binaries in the solar neighborhood (Poveda  et al.
1994). Other authors have searched for CPM companions to nearby
stars (e.g. Lepine \& Shara 2002, Lepine \& Bongiorno 2007, Chaname
\& Gould 2004, Makarov et al. 2008, Scholz et al. 2008) but these
searches have usually set rather stringent criteria for the
acceptable separations and proper motion differences, in order to
eliminate, as far as possible, optical companions.  For our purpose
we set upper limits to the separations of $1.5\arcdeg$,  to the
difference of the proper motions of $\vert\Delta \mu \vert \leq 0.05
\arcsec$~yr$^{-1}$, and to the difference of the position angles of
less than $10\arcdeg$.  These limits would appear at first sight to
be too generous, but we will discuss below some additional criteria
to establish the physical nature of the system
 GJ 282AB - NLTT 18149 in particular.  Here we note only
that the extensively studied system composed of Proxima Centauri and
$\alpha$ Centauri AB is believed to be a physical system (see e.g.
Wertheimer \& Laughlin 2006), despite the fact that Proxima has a
separation of $2.18\arcdeg$ from $\alpha$ Centauri AB, and a
difference in the proper motion of $\vert\Delta \mu \vert =
0.11\arcsec$~yr$^{-1}$.

Our search produced a handful of interesting systems.  In this
paper, we concentrate on GJ 282AB - NLTT 18149, probably the most
remarkable one. For the sake of conciseness we shall refer to this
system as GJ 282 ABC.  The distant companion, NLTT 18149, has a
separation from GJ 282A of $s = 1.09\arcdeg$ and a difference of
proper motion $\vert \Delta \mu \vert = 0.029\arcsec$~yr$^{-1}$. We
will show that all three components of this system have very similar
Hipparcos parallaxes, radial velocities and ages, which, together,
constitute strong evidence in favor of GJ 282ABC being a physical
system.

\section{SEARCH FOR A COMMON PROPER MOTION COMPANION TO GJ 282AB}
\label{sec:cpm}

\begin{table}
\begin{center}
\caption{Some Astrometric and Physical parameters of GJ 282AB and
NLTT 18149}
\begin{tabular}{cccccccccc}
\tableline\tableline
 GJ & NLTT & $\pi^{(1)}$ & $\mu_{\alpha}^{(2)}$ & $\mu_{\delta}^{(2)}$ & $\mu$ & $V_r^{(3)}$ & sep &
sep & Sp$^{(3)}$\\
& & [mas] & arcsec yr$^{-1}$ & arcsec yr$^{-1}$ & arcsec yr$^{-1}$ &
km~s$^{-1}$ & $\arcsec$ & AU \\
\tableline
282A & 18257  &  70.44 & 0.0717$^{(4)}$  &  -0.2761$^{(4)}$  & 0.2852  &  -21.8  &      &        & K2V \\
282B & 18260  &        & 0.0668  &  -0.2862  & 0.2939  &  -22.0  &   58 &    824 & K6.5V\\
282C & 18149  &  69.85 & 0.0363$^{(4)}$  &  -0.2535$^{(4)}$  & 0.2561  &  -22.7  & 3892 &  55733 & M1.5Ve\\
\tableline
\end{tabular}
\tablenotetext{}{(1) Simbad - CDS.  (2) Salim and Gould (2003). (3)
This paper. (4) The proper motions given in van Leeuwen (2007) for components A and C are, in  arcsec yr$^{-1}$, 
$\mu_{\alpha}(A)=0.0699$, $\mu_{\delta}(B)=-0.2786$, $\mu_{\alpha}(C)=0.0374$,  $\mu_{\delta}(C)=-0.2534$. No proper motion for component B is listed.}
\end{center}
\end{table}

Around every primary in our catalogue (Poveda et al. 1994) having
$M_V \leq 9$ and a parallax in Hipparcos (155 primaries), we
searched for common proper motion  companions in the revised NLTT
(Salim \& Gould 2003), looking only at stars within a sphere of 22
pc centered on the Sun, in order to be consistent with the distance
limit of our 1994 catalogue. The magnitude limit ($M_V \leq 9$) was
taken because we showed that our 1994 catalogue is complete up to a
distance of 22 pc for primaries brighter than this magnitude. We
searched for CPM companions within a circle of $1.5\arcdeg$ radius
centered on each primary of our catalogue, differing in proper
motion by less than $0.05 \arcsec$~yr$^{-1}$ and in position angle
by less than $10\arcdeg$ ($\vert\Delta \mu \vert\leq
0.05\arcsec$~yr$^{-1}$, $\Delta \theta \leq 10\arcdeg$.). Note that
the errors in the proper motions of the revised NLTT stars are of
the order of 8 to 10 mas yr$^{-1}$. This would suggest that in order
to find physical pairs one ought to take proper motion differences
of this order. However, such stringent limits would cause us to miss
interesting systems, in which for instance one component (or indeed
both) is an unrecognized astrometric binary, which would introduce a
``spurious'' proper motion difference (due to orbital motion). Also,
taking a proper motion difference of a few mas yr$^{-1}$ will cause
us to miss some of the most interesting, widest, bound systems (e.g.
Proxima Centauri, which has a proper motion difference of 110 mas
~yr$^{-1}$ from Alpha Centauri AB), or systems caught just in the
process of dynamical disintegration (Rodr\'{\i}guez et al. 2005;
G\'omez et al. 2005, 2008; Allen et al. 1974;  Allen et al. 2006;
S\'anchez et al. 2008). Therefore, taking $\vert\Delta \mu \vert
\leq 50$ mas ~yr$^{-1}$ seems to be an adequate compromise to
discover good candidates to wide CPM systems, whose physical
association would need to be confirmed by additional criteria.

Our search revealed several distant companions to binaries.  One of
the most interesting cases is the system GJ 282 AB - NLTT 18149 (GJ
282ABC), having $\vert\Delta \mu \vert = 0.029\arcsec$~yr$^{-1}$,
$\Delta\theta = 6.4\arcdeg$, $s = 1.09\arcdeg$ (see Figure 1).

\section{A COMMON PARALLAX:  ESTIMATE OF THE PROBABILITY OF C BEING AN OPTICAL COMPANION}
Since we take rather generous upper limits for the separations and
proper motion differences it is important to show that the putative
CPM companions we find are not optical.  In the case of GJ 282AB and
NLTT 1814 we proceeded as follows.

The data for  the parallaxes of GJ 282AB and NLTT 18149 given in the
Hipparcos Catalogue are the following:

\begin{eqnarray}
\pi {\rm (GJ \ 282A)} & = 0.07044 \arcsec \pm 0.00094\arcsec\nonumber\\
\pi {\rm (NLTT \ 18149)} & = 0.06985\arcsec \pm
0.00153\arcsec\nonumber .
\end{eqnarray}

\begin{figure}[!t]
  \includegraphics[width=\columnwidth]{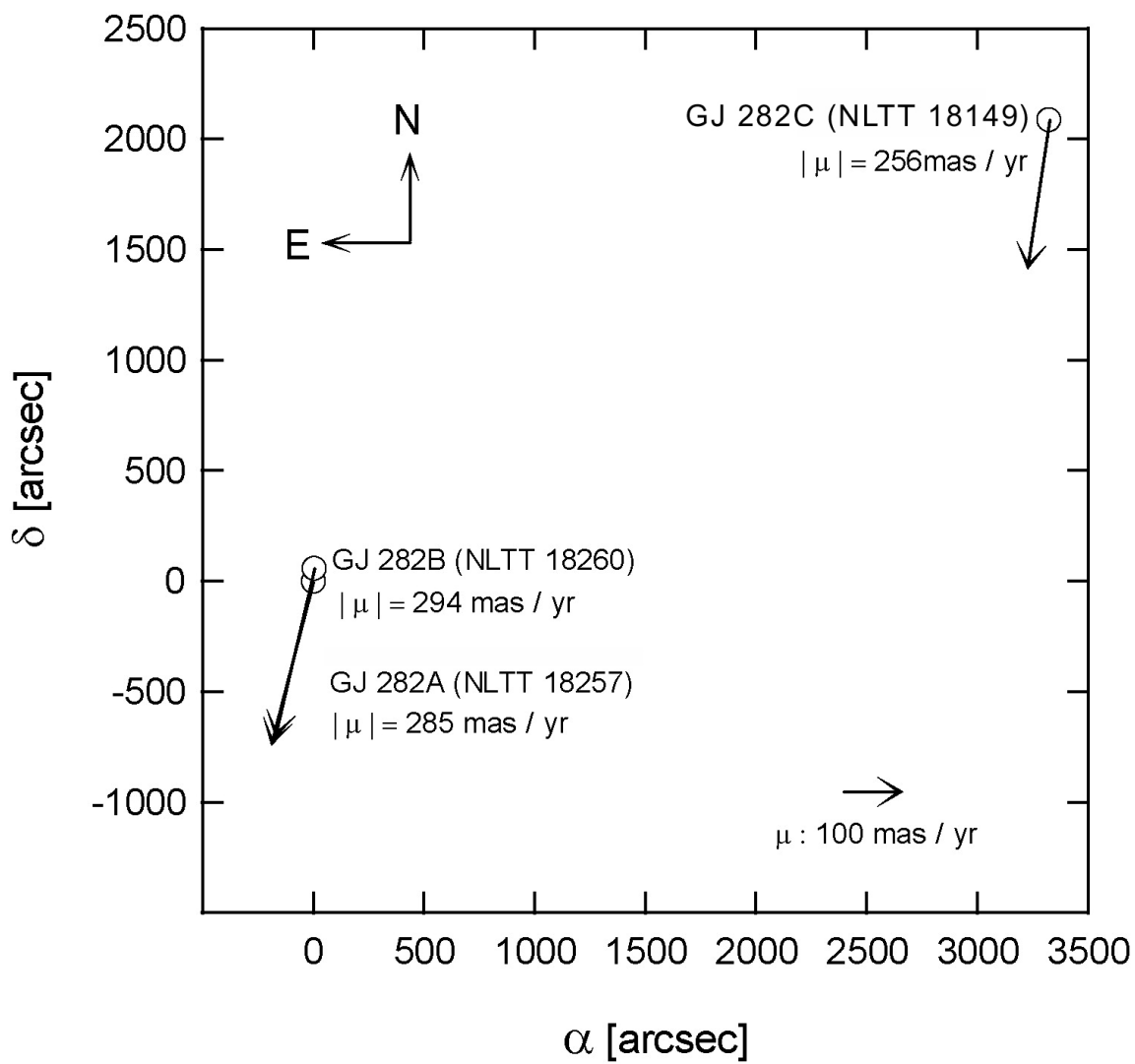}
  \caption{GJ 282 ABC. Relative positions and proper motions.}
  \label{fig:simple}
\end{figure}

These parallaxes are so similar that they allow us to reject the
possibility of NLTT 18149 being an optical companion to GJ 282AB.
Indeed, from the differences of the Hipparcos parallaxes (and their
quoted errors) we estimate that the ``depth'' of the system GJ
282AB-NLTT 18149) is at most 25 000 AU.  This is less than the
projected separation of GJ 282 AB-C, which is 55 285 AU, or about
0.25 pc, so we take the latter as the radius of a sphere and
calculate the expected number of stars contained in this volume,
using the number density determined from the luminosity function by
Reid, Gizis and Hawley (2002), namely, $n = 0.112$ stars per cubic
parsec. The expected number of stars (which could be optical
companions) within this sphere is 0.0072. Next, from the frequency
distribution of $\vert \Delta \mu \vert$ for a representative sample
of the NLTT stars chosen at random we found the probability P($\vert
\Delta \mu \vert \leq 0.05$) = 0.05. The probability of the proper
motion vectors of GJ 282A ($\theta_A =165.44\arcsec$) and NLTT 18149
differing by less than $10\arcdeg$ in position angle is less than
$0.06$, assuming the distribution of position angles to be uniform.
To check this assumption we computed the position angles of the
proper motion vectors of all the rNLTT stars situated in an area of
400 square degrees around GJ 282A (141 stars).  Of these, 30 have
position angles between $155.44\arcdeg$ and $175.44\arcdeg$.
Therefore, our estimate of the probability of the proper motion
vectors differing by less than $10\arcdeg$ has to be increased to
$30/141 = 0.21$. Then, the the probability for one primary from our
catalogue to have an optical NLTT companion satisfying these
restrictions can be calculated to be $p = 0.05 \times 0.21 \times
0.0072 = 7.6 \times 10^{-5}$, and thus the expected number of
opticals (satisfying the above conditions) associated to the 155
systems with $M_V < 9$ in our catalogue is about 0.01. In this
manner, the similarity of the parallaxes and proper motions of the
three components allows us to practically exclude the possibility of
NLTT 18149 being an optical companion to GJ 282AB

\section{A COMMON RADIAL VELOCITY}

\subsection {Observations and spectral classification}

The system GJ 282ABC was observed in 2008 January 19 with the
Echelle spectrograph at the f/7.5 Cassegrain focus of the 2.1 m
telescope of the Observatorio Astr\'onomico Nacional at San Pedro
M\'artir, B.C., M\'exico. The Site3 $1024 \times 1024$ CCD was used
to cover a spectral range from $\lambda$4000 to $\lambda$7100~\AA \
with a spectral resolution of $R \approx 16,000$. Component A was
observed twice, with exposure times of 130 and 300 s each; for
component B we obtained a single 600 s exposure, and for component C
three consecutive exposures, 900 s each. The data reduction was
carried out with the IRAF package\footnote {IRAF is distributed by
the National Optical Observatories, operated by the Association of
Universities for Research in Astronomy, Inc., under cooperative
agreement with the National Science Foundation.}.

\begin{figure}[h]
  \includegraphics[angle=90,width=0.8\columnwidth]{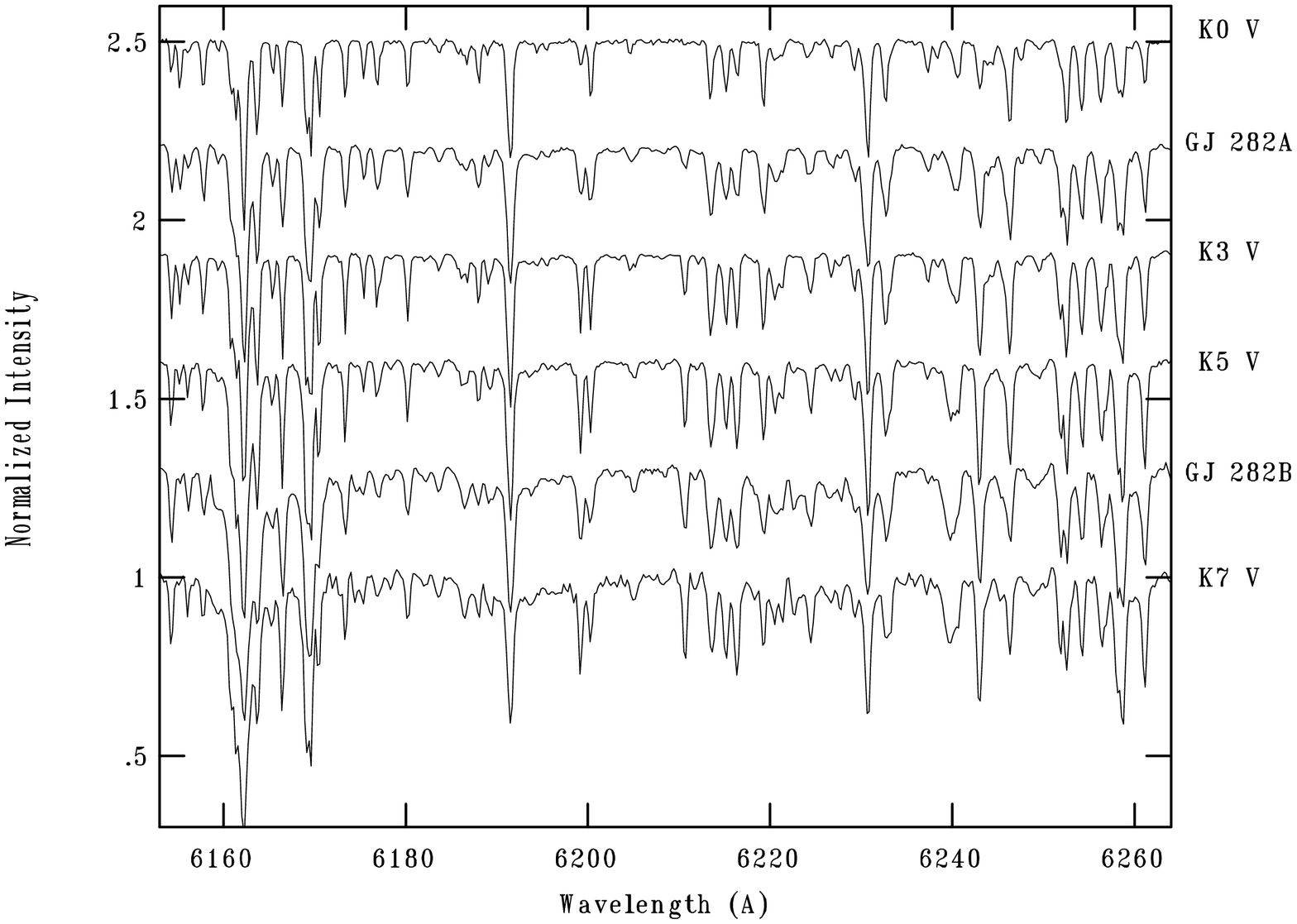}
  \includegraphics[angle=90,width=0.8\columnwidth]{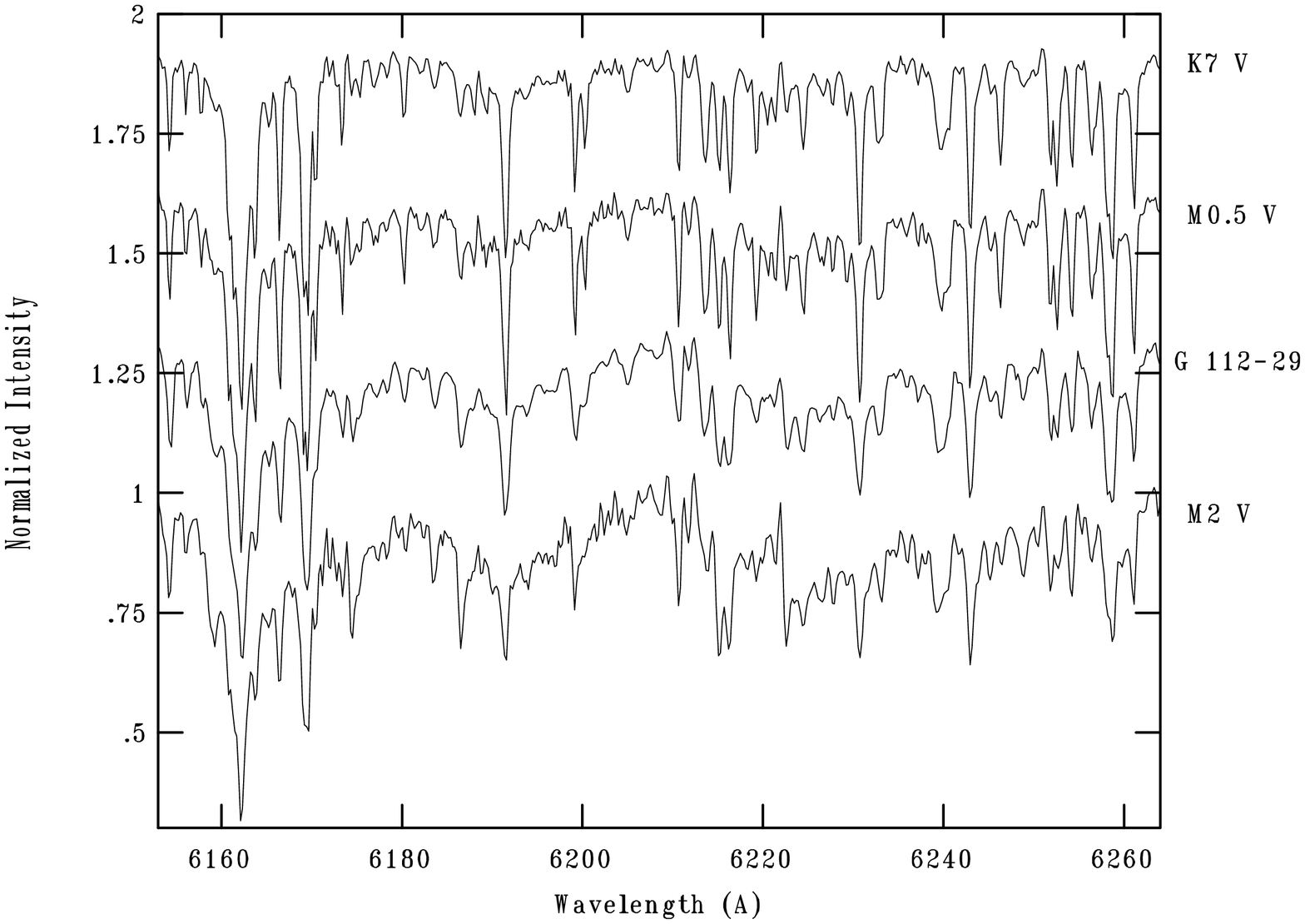}
  \caption{Top: Portion of the normalized spectra of GJ 282A and B compared with the MK
spectral standards used to classify the objects (see Table 2). Each
spectrum has been shifted 0.3 units above the previous one on the
normalized intensity scale and referred to the observer's rest
frame. Several line ratios change rapidly with spectral type,
notably those of Sc\,I $\lambda 6210.7$ \AA \, and V\,I $\lambda
6198.2$ \AA \, relative to their neighboring lines. Bottom: Same as
in Top, but for G 112-29 = GJ 282C. The spectrum of this star is
clearly intermediate between M0.5 and M2, probably closer to the
latter type as inferred from the stronger molecular bands displayed
in other spectral intervals.}
  \label{fig:spectra}
\end{figure}

To classify both components of GJ 282 as well as NLTT 18149 =
G112-29, we compared their spectra with those of the MK standard
stars listed in Table 2.  The latter spectra were obtained in a
previous observing run (Dec. 2006) with the same equipment, but with
a slightly smaller cross-dispersor angle, yielding a spectral
covarage from 3800 to 6850 \AA. With the exception of 61 Cyg A,
which is contained in the list compiled by Morgan \& Keenan (1973),
these are the K0 to M3 main-sequence standards chosen by
Torres-Dodgen \& Weaver (1993), and all of them are included in the
list of standards by Keenan \& McNeil (1976). The comparison was
made in the spectral range spanning from 4800 to 6800 \AA. In
classifying the stars we have excluded the H$\alpha$ and H$\beta$
lines, in order to avoid the possible effects of chromospheric
activity, as implied by the BY Dra--type variability assigned to GJ
282A (=V869 Mon).

The spectrum of GJ 282A is intermediate to those of the K0 and K3-
standards, but closer to the latter. In this spectral range the
intensity of several lines increases rapidly with decreasing
temperature. In Figure 2 we show an illustrative portion of the
spectra of components A and B, together with those of the
corresponding classification standards. Specially useful in this
(and later) spectral ranges is the Sc\,I $\lambda 6210.7$\,\AA \
multiplet 2 line, which is not contaminated with strong lines: it is
very weak at K0\,V then its intensity sharply increases up to the
K7\,V range and continues to increase, but more slowly, beyond. We
conclude GJ 282A is a normal K2\,V star, in accordance with previous
classifications listed in Simbad. However, in the Simbad header for
this object a K2Ve classification is given and attributed to Cenarro
et al. (2007). We certainly find no evidence of line emission in the
entire observed spectral range for this star, and the intensities of
the hydrogen absorption lines look normal for its spectral type.
Cenarro et al. (2007) obtained $T{{\rm eff}} = 4833$ K, $\log\,g =
4.70$ and $[{\rm Fe/H}] = -0.15$ for GJ 282A.

Reid et al. (2004) classified GJ 282B as a K5\,V type star. Several
line ratios in our averaged spectrum of this object are very similar
to those of the K7\,V MK standard. The strength of the above
mentioned Sc\,I line implies a slightly hotter star, but other line
ratios favor a somewhat later type (see Figure 2). With an
uncertainty of 1 spectral subdivision, we estimate that GJ 282B is a
K6.5\,V normal star. The H$\alpha$ line is moderately weaker in GJ
282B than in 61 Cyg B, while the relative intensities of H$\beta$ to
neighboring metalic lines are very similar in both stars, so there
was little, if any, chromospheric activity in the object during its
observation.

The averaged spectrum of NLTT 18149 = G 112-29 is intermediate
between those of GJ 172 (M0.5\,V) and GJ 15A (M2\,V). The TiO and
MgH molecular bands clearly point towards a slightly hotter star
than the M2\,V standard, in accordance with several line ratios
sensitive to temperature in that spectral type range. In addition,
H$\alpha$ presents a weak, double-peaked emission; the peaks are
separated by about $1.4$~\AA \, and the wavelength of the central
trough --that nearly reaches the neighboring continuum level-- is
consistent with that at the rest frame of the star. H$\beta$ is also
weakly (but clearly) in emission, its width being larger than the
instrumental profile, probably due to unresolved structure.
H$\gamma$ and H$\delta$ (with very poor S/N ratio) both appear as
very narow emission lines. Hence, we classify this star as an
M1.5\,Ve star with a small uncertainty (0.5 of a spectral
subdivision). Our classification results are given in Table 1.

\subsection{Radial velocities}

Accurate radial velocities for five of the standards listed in Table
2 have been obtained by Nordstroem et al. (2004) in the CORAvEL
system. In order to check for self consistency or possible small
variations in the velocities published for these five stars, we
first cross-correlated the spectrum of each one with that of the
other four, in the $\lambda\lambda$ 5132-5802 \AA  \ spectral range,
adopting the velocities published by those authors, and using the
IRAF {\it fxcor} task.  The resulting average velocities were in
good agreement with the published ones, except for that obtained for
GJ 105A, which was found to be $23.5 \pm 0.14$ km~s$^{-1}$ (standard
deviation of the velocity as obtained from the four template stars)
slightly but significantly smaller than the published value. When
the exercise was repeated, now substituting the velocity of GJ 105A
with the one we obtained, the average radial velocities of the other
four spectral standards differed by less than 0.3 km~s$^{-1}$ from
the published values, with a $\sigma \leq 0.2$ km~s$^{-1}$. These
five stars constitute the reference system against which we
cross-correlated the spectra of GJ 282 A, B and C, using the radial
velocities listed in the last column of Table 2 and the spectral
range mentioned above. The resulting velocities do not show any
dependence on the spectral type of the template stars, though the
formal errors yielded by the IRAF {\it fxcor} task (between 0.5 and
3.0 km~s$^{-1}$) are smaller for the velocities obtained from
comparison stars with temperatures close to that of the
corresponding object.

The average heliocentric radial velocities so obtained are $-21.8$
km~s$^{-1}$ for GJ 282A, $-22.0$ km~s$^{-1}$ for GJ 282B, and
$-22.7$ km~s$^{-1}$ for NLTT 18149, respectively. The standard
deviation from the mean for the velocities derived from each
template star is less than $0.2$ km~s$^{-1}$ for all three
velocities, whereas their external error is estimated to be $1.5$
km~s$^{-1}$. Of course, the zero point of these measurements is
linked to the CORAVEL system. To our knowledge, of these three stars
only GJ 282A has previously published radial velocities, most
recently by Nordstroem et al. (2004). These authors find a value of
$-18.6 \pm 0.1$ km~s$^{-1}$ for the radial velocity of this star, as
derived from 11 spectra obtained during 13 years, and assign a
probability of $0.62$ for the observed velocity scatter to be due
only to random observational errors. Since this velocity was
measured on the same radial velocity frame as those we obtained, the
difference between their value and ours ($3.2$ km~s$^{-1}$) is
significant and could be due to a small systematic error in the
wavelength calibration of our spectra. However, if real, such an
error would be irrelevant for the purpose of this paper.

In order to improve our estimate of the precision of the velocities
we cross-correlated the two spectral standards with no published
CORAVEL-based radial velocities, with the same five standards as
above. The resulting velocities were $+33.2 \pm 0.6$~km~s$^{-1}$ for
GJ 172 and $+31.5 \pm 1.6$ km~s$^{-1}$ for GJ 752A, which are,
respectively, $2.0$ and $0.9$ km~s$^{-1}$ smaller than the
velocities given by Evans (1967) and listed in Table 2. Hence, we
estimate the precision of the radial velocities here obtained to be,
at most, 2 km~s$^{-1}$.

To directly measure the radial velocity differences between the
three components of GJ 282, we additionally cross-correlated the
spectra of A and C with that of component B;  we obtained $A - B = +
0.1 \pm 1.0$ and $C - B = -0.7 \pm  0.6$ km~s$^{-1}$, in excellent
agreement with the differences obtained from the velocities measured
through the template stars.

We conclude that, when measured on the same reference system, the
three objects have, within errors, the same heliocentric radial
velocity. The slightly different velocity of component C will be
further discussed in Section 6.

\begin{table}[h]
\begin{center}
\caption{Standard stars used in this paper}
\begin{tabular}{lllll}
\tableline\tableline
Name &    HD   &  Sp.T.  &    Pub. $V_r$    &  Adopted $V_r$ \\
     &         &         &    km~s$^{-1}$          &    km~s$^{-1}$       \\
\tableline

$\sigma$ Dra & 185144  &  K0\,V   & $+26.3^{(1)}$ & $+26.3$ \\
GJ 105A      &  16160  &  K3-\,V  & $+25.1^{(1)}$ & $+23.5^{(3)}$\\
61 CygA      & 201091  &  K5\,V   & $-66.5^{(1)}$ & $-66.5$\\
61 CygB      & 201092  &  K7\,V   & $-65.3^{(1)}$ & $-65.3$\\
GJ 172       & 232979  &  M0.5\,V & $+35.2^{(2)}$\\
GJ 15A       & 1326 A  &  M2\,V   & $+11.3^{(1)}$ & $+11.3$\\
GJ 752A      & 180617  &  M3\,V   & $+32.4^{(2)}$\\
\tableline
\end{tabular}
\tablenotetext{}{$^{(1)}$ Evans (1967) $^{(2)}$ Nordstroem et al.
(2004) $^{(3)}$ This paper, see text}
\end{center}
\end{table}

\section{A COMMON AGE}
\label{sec:EPS}

Since at least one of the components of this system is a late-type
H$\alpha$-emission, chromospherically active star, we suspected them
to be X-ray sources. In fact, the three components of this system
turn out to be bright X--ray sources, as observed by the ROSAT
satellite. A relation between the X--ray luminosity $L_x$ and the
age T of low mass stars was obtained by Kunte, Raio and Vahia
(1988).  A similar relation has been recently discussed by Mamajek
and Hillenbrand (2008), who find that coronal activity as measured
by the fractional X--ray luminosity $L_x/L_{{\rm bol}}$ has almost
the same age-inferring capability as does chromospheric activity
measured through the Ca II H and K emission index. We will use both
the Kunte et al. and the Mamajek \& Hillenbrand relations to
estimate the ages of the three stars.

To calculate the ages, X--ray luminosities were taken from the
NEXXUS 2 Database, and bolometric luminosities were calculated from
the observed $V-K$ color index, as given in the Simbad database.
These quantities and the calculated ages according to the two
relations used are listed in Table 3:

>From the straight line fit given in Figure 1 of Kunte, Rao, \& Vahia
(1988) we obtain the relation

\begin{displaymath}
    \log T = 5.63 - 0.654 \log (L_x/L_{{\rm bol}}),
\end{displaymath}

\noindent from which we calculate the ages given in Column 4 of
Table 3.

On the other hand, from equation (A3) in Mamajek \& Hillenbrand
(2008), namely

\begin{displaymath}
    \log T = 1.20 - 2.307 \log (L_x/L_{{\rm bol}}) - 0.1512 [\log (L_x/L_{{\rm bol}})]^2,
\end{displaymath}

\noindent we find the ages given in the last column of Table 3.

\begin{table}[h]
\begin{center}
\caption{Bolometric and X--ray luminosities, and ages}
\begin{tabular}{lcccc}
\tableline\tableline
 Component & $L_x$ & $L_x/L_{{\rm bol}}$ & Age & Age
\\
  & ergs & & yr & yr\\
\tableline

GJ282 A   & $2.45 \times 10^{28}$ & $3.45 \times 10^{-5}$ & $3.5\times 10^{8}$ & $3.0 \times 10^{8}$\\
GJ282 B   & $5.75 \times 10^{27}$ & $1.27 \times 10^{-5}$ & $6.8\times 10^{8}$ & $7.2 \times 10^{8}$\\
NLTT18149 & $1.48 \times 10^{28}$ & $3.86 \times 10^{-5}$ & $3.3\times 10^{8}$ & $2.6 \times 10^{8}$\\
\tableline
\end{tabular}
\end{center}
\end{table}

Table 3 shows that the ages of the three components turn out to be
very similar. If we assume that GJ 282A and B are coeval, then their
age difference is a measure of the uncertainties in the age
determination. Then, the ages of all three stars are equal to within
these uncertainties.

\section{DISCUSSION AND CONCLUSIONS}
\label{sec:conc}

The four properties shared by the system GJ 282 AB-NLTT 18149, to
wit, common proper motions, common parallaxes, common radial
velocities and common ages, constitute  strong arguments in favor of
a physical association of the three stars. Nonetheless, the
difference in their proper motions $\vert\Delta \mu \vert = 0.029
\arcsec$/year is large enough (compared to the error) to be a cause
for concern.

There are several possible explanations for this difference.
Physically bound systems with large angular separations may have
slightly different $\mu_\alpha$, $\mu_\delta$, $v_r$ simply due to
projection effects. Orbital motion of GJ 282AB may also cause such
differences. We examine each of these effects in turn. To assess the
importance of projection effects we assume that the space motion of
G112-29 is identical to that of GJ282A.  Using the standard formulas
(e.g. Smart 1962, p.16) we calculate the $\Delta\mu_\alpha$,
$\Delta\mu_\delta$, and $\Delta v_r$ resulting from the different
positions and distances of GJ 282A and G112-29. We obtain
$\Delta\mu_\alpha = 0.007\arcsec$~yr$^{-1}$, $\Delta\mu_\delta =
0.004\arcsec$~yr$^{-1}$ and $\Delta v_r = 0.23$ km~s$^{-1}$.  These
values are much smaller than the observed differences, and hence we
conclude that the latter are not due solely to projection effects.

To estimate the importance of orbital motion of the pair GJ 282AB we
need the masses of both components. These were calculated from their
photometry and the mass-luminosity relation of Reid et al. (2002),
and turn ut to be $M_A  = 0.7~M_\odot$, $M_B = 0.5~M_\odot$.
Assuming the orbit to be circular and perpendicular to the plane of
the sky, we obtain a maximum contribution to the radial velocity of
$\Delta v_r = 0.49$ km~s$^{-1}$. The maximum contribution to the
tangential velocity is also $0.49$ km~s$^{-1}$, which corresponds to
a maximum proper motion of $\vert\Delta\mu\vert = 0.007\arcsec$
yr$^{-1}$, much smaller than the observed  $\vert\Delta\mu\vert =
0.029\arcsec$ yr$^{-1}$. Conversely, if the orbit is in the plane of
the sky, we obtain a maximum contribution to the proper motion of
$\vert\Delta\mu\vert = 0.007\arcsec$ yr$^{-1}$, again much smaller
than the observed $\Delta\mu$. In this case, orbital motion
contributes nothing to the observed $\Delta v_r$. We conclude that
the orbital motion of GJ 282AB could marginally account for the
observed $\Delta v_r$, but even in the extreme case of an orbit
wholly in the plane of the sky cannot account for the observed
$\vert\Delta\mu\vert$. If the orbit is eccentric, we have to
multiply these values by at most a factor of 1.4, but the conclusion
remains unchanged.

Having shown that both projection effects and orbital motion cannot
explain the discrepancy of $\vert\Delta\mu\vert = 0.029\arcsec$
yr$^{-1}$ we suggest that the system GJ 282 AB - NLTT 18149 is in
the process of dynamical disintegration. It would not be a unique
case. There are at least two other multiple systems that appear to
be in a similar state. The system $\theta_{1}$ Orionis ABCD - E
(Allen, Poveda \& Worley 1974; Allen, Poveda \&
Hern\'andez-Alc\'antara 2006; S\'{a}nchez et al. 2008) has been
shown to be in the process of ejecting component E, and  the system
composed of BN - I - n (Rodr\'\i guez et al. 2005, G\'omez et al.
2006) is also disintegrating.  Thus, GJ 282AB-NLTT 18149 could be
another member of the interesting new class of systems caught in the
process of gravitational disintegration. A rough calculation shows that component C would have been ejected about 60,000 years ago. This value is much smaller than the estimated ages of the stars, which implies that the dynamical evolution of the hypothetical bound triple ABC proceeded slowly during most of its lifetime, before C finally escaped.

{\it Acknowledgments.} AP is gratefull to the Direcci\'{o}n General
de Servicios de C\'{o}mputo Acad\'{e}mico (DGSCA - UNAM) for the
facilities granted. This paper used the Simbad database and IRAF
facilities. We are grateful to the anonymous referee for comments
that improved this paper.


\begin{thebibliography}


\bibitem{allen1} Allen, C.; Poveda, A.; Worley, C. E. 1974, RevMexAA, 1, 101

\bibitem{allen2} Allen, C.; Poveda, A.; Hern\'andez-Alcantara, A. 2006, RevMexAA (SC), 25, 13

\bibitem{cenarro} Cenarro A.J.; Peletier, R.F.; S\'anchez-Bl\'azquez, P.; Selam, S.O.; Toloba, E.; Cardiel,
N.; Falc\'on-Barroso, J.; Gorgas, J.; Jim\'enez-Vicente, J.;
Vazdekis, A. 2007, MNRAS, 374, 664

\bibitem{Chaname} Chaname, J.; Gould, A. 2004, ApJ, 601, 289

\bibitem{evans} Evans, D.S. 1967, IAU Symp., 30, 57

\bibitem{Gomez1} G\'omez, L.; Rodr\'{\i}guez, L.F.; Loinard, L.; Lizano, S.;
    Poveda, A.; Allen, C.  2005, ApJ, 635, 1166

\bibitem{Gomez2} G\'omez, L.; Rodr\'{\i}guez, L.F.; Loinard, L.; Lizano, S.;
    Allen, C.; Poveda, A.; Menten, K.M. 2008, ApJ, 685, 333

\bibitem{keenan} Keenan, P.C. \& McNeil, R.C. 1976, {\it An Atlas of the Spectra of
the Cooler Stars (Colombus, Ohio State University Press)}

\bibitem{Kunte} Kunte, P.K.; Rao, A.R.;  Vahia, M. 1988, Ap\&SS, 143, 207

\bibitem{Lepine1} Lepine, S.;  Bongiorno, B. 2007, AJ, 133, 889

\bibitem{Lepine2} Lepine, S.;  Shara, M.M.; Rich, R.M. 2002, AJ, 123,3434

\bibitem{Makarov} Makarov, V.V.; Zacharias, N.; Hennessy, G.H. 2008, ApJ, 687, 566

\bibitem{Mamajek} Mamajek, E.E.; Hillenbrand, L.A. 2008, ApJ, 687,1264

\bibitem{Morgan} Morgan, W.W. \& Keenan, P.C. 1973, Ann. Rev. A\&A, 11, 29

\bibitem{Nexxus}NEXXUS 2 The Database for Nearby X-Ray and extreme UV emitting Stars
http://www.hs.uni-hamburg.de/DE/For/Gal/Xgroup/nexxus/nexxus.html

\bibitem{Norstroem} Nordstroem, B. et al. 2004, Astron. \& Astrophys., 418, 989

\bibitem{Poveda} Poveda, A.; Herrera, M.A.; Allen, C.; Cordero, G.; Lavalley, C. 1994, RevMexAA, 28, 43

\bibitem{Reid} Reid, I.N.; Gizis, J.E.;  Hawley, S.L. 2002, AJ, 124, 2721

\bibitem{Reid} Reid, I.N. et al. 2004, AJ, 128, 463

\bibitem{Rodriguez} Rodr\'{\i}guez, L.F.; Poveda, A.; Lizano, S.; Allen, C. 2005, ApJ, 627, 65

\bibitem{Salim} Salim, S.; Gould, A. 2003, ApJ, 582, 1011


\bibitem{Sanchez} S\'anchez, L.J.; Ruelas-Mayorga, A.; Allen, C.; Poveda, A. 2008,
RevMexAA(SC), 34, 10

\bibitem{Scholz} Scholz, R.D.; Kharchenko, N.V.; Lodieu, N.; McCaughrean 2008, A\&A, 487, 595

\bibitem{Smart} Smart, W.M. 1968, ''Stellar Kinematics'' (New York,
Y.Y, Wiley and Sons Inc.), p. 16

\bibitem{torres-d} Torres-Dodgen, A.V. \& Weaver, Wm.B. 1993, PASP, 105, 693

\bibitem{vanLeeuwen} van Leeuwen, F. 2007, A\&A, 474, 653

\bibitem{Wertheimer} Wertheimer, J.G.; Laughlin, G. 2006, AJ, 132, 1995

\end{thebibliography}
\end{document}